\documentstyle[eqsecnum,aps]{revtex}
\begin{document}
\draft
\baselineskip= 12 pt
\title{\bf
Nambu-Goldstone mechanism at finite temperature in the imaginary-time and real-time 
formalism
\\
}
\author{
{\bf
Bang-Rong Zhou}\thanks{Electronic mailing address: zhoubr@163bj.com}
}
\address{  
China Center of Advanced Science and Technology (World Laboratory)
P.O. Box 8730, Beijing 100080, China \\
and \\
Department of Physics, Graduate School, Academia Sinica, Beijing 100039, China\thanks{Mailing address.} \\
}
\date{}
\maketitle 
\begin{abstract}
In the imaginay-time formalism of thermal field theory, and also in the real-time
formalism but by means of some redefined physical propagators for scalar bound states
by diagonalization of four-point function matrices, we reexamine the
Nambu-Goldstone mechanism of electroweak symmetry breaking in a one-generation
fermion condensate scheme, based on the Schwinger-Dyson equation in the fermion
bubble diagram approximation, and compare the obtained results. We have reached the
conclusion that in both the formalisms, the Goldstone theorem of spontaneous
electroweak symmetry breaking is rigorously true for the case of mass-degenerate 
two flavors of fermions and only approximately valid at low energy scales for the
mass-nondegenerate case, in spite of existence of some difference between the two 
formalisms in the imaginary parts of the denominators of the propagators for scalar 
bound states. When the two flavors of fermions have unequal nonzero masses, the 
induced possible fluctuation effect for the Higgs particle  is negligible if the 
momentum cut-off in the zero temperature loops is large enough.  All the results 
show  physical equivalence of the two formalisms in the present discussed problems.
\end{abstract}
\section{Introduction}
In the research on spontaneous symmetry breaking at finite temperature [1-6],
besides the central problem of phase transition and critical behavior of a
system, theoretical exploration of the Nambu-Goldstone mechanism [7-8]  at
finite temperature is certainly quite significant  for deeper understanding
of spontaneous symmetry breaking [9-11]. When symmetry breaking is induced
dynamically by fermion condensates,  the Nambu-Jona-Lasinio (NJL) model with
four-fermion interactions may be a simple and physically clear laboratory for
this research [8].  The key point of such research lies in verifying existence
of the Nambu-Goldstone bosons as products of spontaneous symmetry breaking, i.e.
determining the masses of relevant scalar and pseudoscalar bound states
consisting of fermions and antifermions. For the sake of examining the
mass-difference effect of constituent fermions in a bound state, we prefer
the Schwinger-Dyson equation approach of the Green functions  to the auxialiary
scalar field method which was extensively used in research on models of
the NJL form [6]. \\
\indent Based on above strategy, we have researched the Nambu-Goldstone
mechanism at finite temperature in the real-time formalism of thermal field
theory in two models of the NJL form [10-11]. It was shown that the Goldstone
theorem is true rigorously if the constituent fermions of a bound state have
the same masses, otherwise it is only valid approximately at low energy scales.
The mass difference between the fermions and the antifermions in a bound state
could lead to  that the Higgs boson has doubled masses and some would-be 
Goldstone bosons will no longer be massless rigorously. However, it should be 
indicated that in obtaining the above results we have used a special definition of 
the propagators for relative scalar bound states. Although the effects
induced by the mass difference of the fermions are negligible when the momentum
cut-off is very large, we still want to know if the above results represent
general conclusions of a thermal field theory, or some of them is only due to the
use of the special definition of the propagators for  scalar bound states in the 
real-time formalism  there.  \\
\indent To clarify this problem, in this paper, based on the same strategy as above,
we will reexamine the Nambu-Goldstone mechanism at finite temperature first in the 
imaginary-time formalism of thermal field theory  and then also in the real-time 
formalism but by means of some redefined physical propagators for scalar bound states [12]. We will again take the one-generation fermion condensate  scheme of electroweak 
symmetry breaking and work by the Schwinger-Dyson equation  in the fermion bubble 
diagram approximation. \\
\indent The paper is arranged as follows. In Sec. II we give the Lagrangian of
the model and the gap equation at finite temperature in the imaginary-time
formalism. In Sec. III we will first calculate the Matsubara propagator for
scalar bound state, then continue analytically it for the energy of the bound state
from discrete frequency to physical values and determine the physical mass of
the scalar bound state. In Secs. IV and V. the same procedure will be applied
to pseudoscalar and charged-scalar bound states. In Sec. VI we will derive the
redefined physical propagators for scalar bound states in the real-time formalism
and compare the results obtained in the two formalisms.  Finally in Sec. VII our 
conclusions follow.
\section{Gap equation in imaginary-time formalism}
In the one-generation fermion condensate scheme of electroweak symmetry
breaking, the one generation of $Q$ fermions form a
${\rm SU}_L(2)\times {\rm U}_Y(1)$  doublet $(U,D)$ and are assigned
in the representation $R$ of the color group ${\rm SU}_c(3)$ with the dimension
$d_Q(R)$.   The symmetry breaking is induced by the effective four-fermion
Lagrangian among the $Q$ fermions below some high momentum scale $\Lambda$ [12]
\begin{equation}
{\cal L}_{4F}= {\cal L}^S_{4F} + {\cal L}^P_{4F} + {\cal L}^C_{4F},
\end{equation}
where the neutral scalar couplings
\begin{equation}
{\cal L}^S_{4F}=\frac{1}{4}\sum_{Q,Q'}
                g_{Q'Q}{(\bar{Q'}Q')}{(\bar{Q}Q)}
\end{equation}
\noindent with
\begin{equation}
               g_{Q'Q}=g^{1/2}_{Q'Q'}
                              g^{1/2}_{QQ}, \ \ Q,Q'=U,D,
\end{equation}
the neutral pseudoscalar couplings
\begin{equation}
{\cal L}^P_{4F}=\frac{1}{4}\sum_{Q,Q'}
                g'_{Q'Q}{(\bar{Q'}i\gamma_5Q')}
                                        {\bar{Q}i\gamma_5Q)}
\end{equation}
\noindent with
\begin{equation}
           g'_{Q'Q}={(-1)}^{I^3_{Q'}-I^3_Q}g_{Q'Q}, \ \ Q,Q'=U,D,
\end{equation}
and the denotation $I^3_Q$ being the third component of the weak isospin of
the $Q$ fermions, and the charged scalar couplings
\begin{equation}
{\cal L}_{4F}^C=\frac{G}{2}
{(\bar{D}\Gamma^+U)}{(\bar{U}\Gamma^-D)}
\end{equation}
\noindent with
\[
\Gamma^{\pm}=\frac{1}{\sqrt{2}}[\cos\varphi -\sin\varphi \pm(\cos\varphi +
\sin\varphi )\gamma_5],
\]
\begin{equation}
G=g_{UU}+g_{DD}, \ \ \ \ \cos^2\varphi =g_{UU}/G, \ \ \ \ \sin^2\varphi =g_{DD}/G.
\end{equation}
\noindent ${\cal L}_{4F}$ can also be expressed by [13]
\begin{equation}
{\cal L}_{4F}=\frac{G}{4}\left[
{(\phi^0_S)}^2+{(\phi^0_P)}^2+2\phi^+\phi^-\right],
\end{equation}
where
\[
\phi^0_S=\cos\varphi {(\bar{U}U)}+\sin\varphi{(\bar{D}D)},
\]
\[
\phi^0_P=\cos\varphi {(\bar{U}i\gamma_5U)}-
         \sin\varphi{(\bar{D}i\gamma_5D)},
\]
\begin{equation}
\phi^-={(\bar{U}\Gamma^-D)}, \ \ \phi^+={(\bar{D}\Gamma^+U)}
\end{equation}
are, respectively, the configurations of the physical neutral scalar, neutral
pseudoscalar, and charged scalar bound states.
In the imaginary-time (Euclidean) field theory, we will use the conventional
time-space coordinate $(\tau =it, \stackrel{\rightharpoonup}{x})$, the
four-momentum
\begin{equation}
\bar{p}=({\bar{p}}^0,{\bar{p}}^i)=(ip^0, p^i)
\end{equation}
\noindent and the $\gamma$-matrices in spinor space
\begin{equation}
{\bar{\gamma}}^0=i\gamma^0, \ \ \ {\bar{\gamma}}^i=\gamma^i, \ \ \
\gamma_5=i\gamma^0 \gamma^1 \gamma^2 \gamma^3
        ={\bar{\gamma}}^0 {\bar{\gamma}}^1 {\bar{\gamma}}^2 {\bar{\gamma}}^3
\end{equation}
\noindent which submit to the anticommulation relations
\begin{equation}
\{ {\bar{\gamma}}^{\mu},{\bar{\gamma}}^{\nu} \}=-2\delta^{\mu \nu}, \ \ \
\{ {\bar{\gamma}}^{\mu},\gamma_5 \}= 0,
\end{equation}
\noindent where $\gamma^{\mu}(\mu=0,1,2,3)$ are the ordinary $\gamma$-matrices
in the real-time (Minkowski) field theory. In this way, the propagator in the
momentum space for the $Q$ fermion with mass $m_Q$ and chemical potential
$\mu_Q$   will  be expressed by
\begin{equation}
\frac{m_Q-{\bar{\not l}}_Q}{(\omega_n+i\mu_Q)^2+
{\stackrel{\rightharpoonup}{l}}^2 + m_Q^2}
=S_Q(-i\omega_n +\mu_Q, \stackrel{\rightharpoonup}{l}), \ \ \
\omega_n=\frac{(2n+1)\pi}{T}, \ \ \
{\bar{\not l}}_Q = {\bar{\gamma}}^{\mu}{\bar{l}}_Q^{\mu}, \ \ \
{\bar{l}}_Q^{\mu}=(\omega_n+i\mu_Q, \stackrel{\rightharpoonup}{l})
\end{equation}
\noindent and the Feynman rule, for example, corresponding to the four-fermion
couplings in ${\cal L}_{4F}^S$, will be $g_{Q'Q}/2$. \\
\indent The derivation of the gap equation is similar to that made in the
real-time formalism [11], the main change is to replace the integral of the loop
energy by the sum of Matsubara frequency. Therefore when assuming the thermal
expectation value $\sum_{Q=U,D}g_{QQ}{\langle{\bar{Q}Q}\rangle}_{T} \neq 0$
we will obtain  the mass of the $Q$ fermion
\begin{equation}
m_Q(T, \mu)\equiv m_Q=-\frac{1}{2}g^{1/2}_{QQ}\sum_{Q'=U,D}g^{1/2}_{Q'Q'}
                      {\langle\bar{Q'}Q'\rangle}_T ,
\end{equation}
which will lead to the relation
\begin{equation}
m_Q/g^{1/2}_{QQ}=m_{Q'}/g^{1/2}_{Q'Q'}
\end{equation}
and the gap equation at finite temperature $T$,
\begin{equation}
1=\sum_{Q=U,D}g_{QQ}I_Q,
\end{equation}
with
\begin{eqnarray}
I_Q=-\frac{1}{2m_Q}{\langle{\bar{Q}Q}\rangle}_T
  &=&\frac{d_Q(R)}{2m_Q}\int \frac{d^3l}{{(2\pi)}^3}
   T\sum_{n=-\infty}^{\infty}
   \frac{{\rm tr}(m_Q-{\bar{\not l}}_Q)}{{(\omega_n+i\mu_Q)}^2+
{\stackrel{\rightharpoonup}{l}}^2 + m_Q^2}\nonumber
                 \\
  &=&2d_Q(R)\int \frac{d^3l}{{(2\pi)}^3}T\sum_{n=-\infty}^{\infty}
     \frac{1}{ {(\omega_n+i\mu_Q)}^2+\omega_{Ql}^2 },
\end{eqnarray}
\begin{equation}
\omega_{Ql}^2={\stackrel{\rightharpoonup}{l}}^2 + m_Q^2.
\end{equation}
To find the frequency sum here and later, we define the Fourier transform
[14]
\begin{equation}
\frac{1}{ {(\omega_n+i\mu_Q)}^2+\omega_{Ql}^2 }
=\int_{0}^{\beta}d\tau e^{-i\omega_n\tau}\tilde{\Delta}(\tau,\omega_{Ql},
\mu_Q)
\end{equation}
with $\beta=1/T$ and the inverse formula
\begin{equation}
\tilde{\Delta}(\tau,\omega_{Ql},\mu_Q)=T\sum_{n=-\infty}^{\infty}
e^{i\omega_n\tau}\frac{1}{ {(\omega_n+i\mu_Q)}^2+\omega_{Ql}^2 }.
\end{equation}
The $\tilde{\Delta}(\tau,\omega_{Ql},\mu_Q)$ in Eq. (2.20) obeys the
antiperiodicity condition
\begin{equation}
\tilde{\Delta}(\tau,\omega_{Ql},\mu_Q) =
-\tilde{\Delta}(\tau-\beta,\omega_{Ql},\mu_Q)
\end{equation}
and can be calculated by the formula
\[\tilde{\Delta}(\tau,\omega_{Ql},\mu_Q)=
\frac{i}{4\pi}\int_{C_1\cup C_2}dz f(z) \tan \frac{\beta}{2}(z-i\mu_Q),\]
\begin{equation}
f(z)=e^{i(z-i\mu_Q)\tau}\frac{1}{z^2+\omega^2_{Ql}},
\end{equation}
where $C_1$ and $C_2$ represent the integral paths $-\infty +
i(\mu_Q-\varepsilon) \longrightarrow +\infty +i(\mu_Q-\varepsilon)$ and
$+\infty +i(\mu_Q+\varepsilon) \longrightarrow -\infty +i(\mu_Q+\varepsilon)$
respectively in complex $z$ plane.  The result is
\begin{equation}
\tilde{\Delta}(\tau,\omega_{Ql},\mu_Q)=\frac{1}{2\omega_{Ql}}
\{[1-n(\omega_{Ql}-\mu_Q)]e^{-(\omega_{Ql}-\mu_Q)\tau}-
     n(\omega_{Ql}+\mu_Q)e^{(\omega_{Ql}+\mu_Q)\tau} \},
\end{equation}
where the denotations
\begin{equation}
n(\omega_{Ql}\pm \mu_Q) =1/\left[e^{\beta(\omega_{Ql}\pm \mu_Q)}+1\right]
\end{equation}
has been used.  Obviously, it can be obtained from Eqs. (2.20)  and (2.23)
that
\begin{eqnarray}
T\sum_{n=-\infty}^{\infty}\frac{1}{ {(\omega_n+i\mu_Q)}^2+\omega_{Ql}^2 }
&=&\tilde{\Delta}(\tau=0,\omega_{Ql},\mu_Q) \nonumber \\
&=&\frac{1}{2\omega_{Ql}}[1-n(\omega_{Ql}-\mu_Q)-n(\omega_{Ql}+\mu_Q)]
\nonumber \\
&=&\int_{-\infty}^{\infty}\frac{dl^0}{2\pi}\left[
   \frac{i}{{l^0}^2-\omega_{Ql}^2+i\varepsilon}-2\pi
    \delta ({l^0}^2-\omega^2_{Ql})\sin^2\theta(l^0, \mu_Q)
  \right],
\end{eqnarray}
with the definition
\begin{equation}
\sin^2\theta(l^0,\mu_Q)=\frac{\theta(l^0)}{\exp[\beta (l^0-\mu_Q)]+1}
                        +\frac{\theta(-l^0)}{\exp[\beta (-l^0+\mu_Q)]+1}.
\end{equation}
Substituting Eq. (2.25) into Eq. (2.17), we can express the gap equation (2.16)
by
\begin{equation}
1=\sum_{Q=U,D}g_{QQ}2d_Q(R)\int \frac{d^4l}{{(2\pi)}^4}
  \left[
   \frac{i}{l^2-m_Q^2+i\varepsilon}-2\pi  \delta (l^2-m^2_Q)
   \sin^2\theta(l^0, \mu_Q)\right],
\end{equation}
which is precisely the expression of the gap equation in the real-time
formalism [11].  It should be indicated that such identity of the gap equation
in the two formalism depend on the fact that the equation comes from  the
two-point Green function and is determined by  the loops bounded by a single
fermion propagator.  Equation (2.16) or (2.27), as has been shown [15], could be
satisfied  only  at the temperature $T$ below the electroweak symmetry
restoration temperature $T_c$.  Therefore, in the following discussions we will
always assume $T<T_c$ and the gap equation being obeyed.
\section{Scalar bound state in the imaginary-time formalism}
Since a bound state is formed by the four-fermion interactions, its propagator must 
correspond to a four-point amputated Green function.  Thus in the imaginary-time 
formalism, the propagator for a scalar bound state can be
calculated by means of the four-point amputated functions
$\Gamma_S^{Q'Q}(-i\Omega_m, \stackrel{\rightharpoonup}{p})$ for the transition
from $(\bar{Q}Q)$ to $(\bar{Q'}Q')$, where
\begin{equation}
\Omega_m=\frac{2\pi m}{T}, \ \ \ m=0, \pm 1, \pm 2,\cdots
\end{equation}
represent the Matsubara frequency of the scalar bound state and
$\stackrel{\rightharpoonup}{p}$ is its three dimension momentum. Based on the
scalar four-fermion couplings ${\cal L}^S_{4F}$ in Eq. (2.2), in the fermion
bubble diagram approximation,
$\Gamma_S^{Q'Q}(-i\Omega_m, \stackrel{\rightharpoonup}{p})$ must obey the 
algebraic
equations
\begin{equation}
\Gamma_S^{Q'Q''}(-i\Omega_m, \stackrel{\rightharpoonup}{p})[
\delta_{Q''Q}-N_{Q''}(-i\Omega_m, \stackrel{\rightharpoonup}{p}) g_{Q''Q}]
=g_{Q'Q}/2,
\end{equation}
where $2N_{Q}(-i\Omega_m, \stackrel{\rightharpoonup}{p})$ is the contribution of
the $Q$ fermion loop with scalar coupling vertices and
\begin{equation}
N_{Q}(-i\Omega_m, \stackrel{\rightharpoonup}{p})
=-\frac{d_Q(R)}{2}\int \frac{d^3l}{(2\pi)^3}
 T\sum_{n=-\infty}^{\infty}{\rm tr}
 S_Q(-i\omega_n+\mu_Q, \stackrel{\rightharpoonup}{l})
 S_Q(-i\omega_n-i\Omega_m+\mu_Q, \stackrel{\rightharpoonup}{l} +
                                 \stackrel{\rightharpoonup}{p}).
\end{equation}
Equations (3.2) have the solutions
\begin{equation}
\Gamma_S^{Q'Q}(-i\Omega_m, \stackrel{\rightharpoonup}{p})
=\frac{g_{Q'Q}}{2\Delta(-i\Omega_m, \stackrel{\rightharpoonup}{p})},
\end{equation}
where
\begin{equation}
\Delta(-i\Omega_m, \stackrel{\rightharpoonup}{p})
=1-\sum_{Q=U,D}g_{QQ}N_{Q}(-i\Omega_m, \stackrel{\rightharpoonup}{p}).
\end{equation}
The propagator for the scalar bound state $\phi_S^0$ shown in Eq. (2.9) becomes
\begin{equation}
\Gamma_{I}^{\phi_S^0}(-i\Omega_m, \stackrel{\rightharpoonup}{p})
=G/2\Delta(-i\Omega_m, \stackrel{\rightharpoonup}{p}).
\end{equation}
By means of Eqs. (2.13), (2.12)  and (3.3), we find
\begin{equation}
N_Q(-i\Omega_m, \stackrel{\rightharpoonup}{p}) = I_Q+
2d_Q(R)\int \frac{d^3l}{(2\pi )^3}T\sum_{n}
\frac{-(\Omega_m^2+{\stackrel{\rightharpoonup}{p}}^2)-2m_Q^2
      -(\omega_n+i\mu_Q)\Omega_m-\stackrel{\rightharpoonup}{l}\cdot
       \stackrel{\rightharpoonup}{p}}
     {[(\omega_n+i\mu_Q)^2+{\stackrel{\rightharpoonup}{l}}^2+m_Q^2]
      [(\omega_n+\Omega_m+i\mu_Q)^2+{(\stackrel{\rightharpoonup}{l}+
                                      \stackrel{\rightharpoonup}{p})}^2+m_Q^2]}.
\end{equation}
From Lorentz invariance, in the imaginary-time formalism,
$\Gamma_{I}^{\phi_S^0}(-i\Omega_m, \stackrel{\rightharpoonup}{p})$ should be a
 function
of $-(\Omega_m^2+{\stackrel{\rightharpoonup}{p}}^2)$, thus must obey the
constraint
\begin{equation}
\Gamma_{I}^{\phi_S^0}(-i\Omega_m, \stackrel{\rightharpoonup}{p})
=\Gamma_{I}^{\phi_S^0}(+i\Omega_m, -\stackrel{\rightharpoonup}{p})
=\Gamma_{I}^{\phi_S^0}(-i\Omega_{-m}, -\stackrel{\rightharpoonup}{p}).
\end{equation}
From Eqs. (3.6) and (3.5),  the same constraint on
$N_Q(-i\Omega_m, \stackrel{\rightharpoonup}{p})$  is implied and this will lead
to the equality
\begin{eqnarray}
\lefteqn{\int d^3lT\sum_{n}\frac{-(\omega_n+i\mu_Q)\Omega_m-
                \stackrel{\rightharpoonup}{l}\cdot\stackrel{\rightharpoonup}{p}}
     {[(\omega_n+i\mu_Q)^2+{\stackrel{\rightharpoonup}{l}}^2+m_Q^2]
      [(\omega_n+\Omega_m+i\mu_Q)^2+
      {(\stackrel{\rightharpoonup}{l}+\stackrel{\rightharpoonup}{p})}^2+m_Q^2]}
}  \nonumber \\
& & =\int d^3lT\sum_{n}\frac{(\Omega_m^2+{\stackrel{\rightharpoonup}{p}}^2)/2}
     {[(\omega_n+i\mu_Q)^2+{\stackrel{\rightharpoonup}{l}}^2+m_Q^2]
      [(\omega_n+\Omega_m+i\mu_Q)^2+
      {(\stackrel{\rightharpoonup}{l}+\stackrel{\rightharpoonup}{p})}^2+m_Q^2]}.
\end{eqnarray}
As a result, we obtain
\begin{equation}
N_Q(-i\Omega_m, \stackrel{\rightharpoonup}{p})
=I_Q-\frac{1}{2}(\Omega_m^2+{\stackrel{\rightharpoonup}{p}}^2+4m_Q^2)
                K_Q^T(-i\Omega_m, \stackrel{\rightharpoonup}{p}),
\end{equation}
where
\begin{equation}
K_Q^T(-i\Omega_m, \stackrel{\rightharpoonup}{p})
=2d_Q(R)\int \frac{d^3l}{(2\pi)^3}
A_Q(-i\Omega_m, \stackrel{\rightharpoonup}{p},\stackrel{\rightharpoonup}{l}),
\end{equation}
\begin{equation}
A_Q(-i\Omega_m, \stackrel{\rightharpoonup}{p},\stackrel{\rightharpoonup}{l})
=T\sum_{n}\frac{1}{(\omega_n+i\mu_Q)^2+\omega^2_{Ql}}
          \frac{1}{(\omega_n+\Omega_m+i\mu_Q)^2+\omega^2_{Ql+p}},
\end{equation}
\begin{equation}
\omega^2_{Ql+p}=
{(\stackrel{\rightharpoonup}{l}+\stackrel{\rightharpoonup}{p})}^2+m_Q^2.
\end{equation}
By means of Eqs. (2.19), (2.23) and the formula
\begin{equation}
T\sum_{n}e^{i\omega_n(\tau-\tau')}=\delta(\tau-\tau'),
\end{equation}
we can find out the frequency sum (3.12) and obtain
\begin{eqnarray}
A_Q(-i\Omega_m, \stackrel{\rightharpoonup}{p},\stackrel{\rightharpoonup}{l})
&=&\int_{0}^{\beta}d\tau e^{-i\Omega_m \tau}
         \tilde{\Delta}(\tau, \omega_{Ql}, -\mu_Q)
         \tilde{\Delta}(\tau, \omega_{Ql+p}, \mu_Q) \nonumber \\
&=&\frac{1}{4\omega_{Ql}\omega_{Ql+p}}\left\{
\frac{1-n(\omega_{Ql}+\mu_Q)-n(\omega_{Ql+p}-\mu_Q)}
     {i\Omega_m+\omega_{Ql}+\omega_{Ql+p}}
+\frac{n(\omega_{Ql}+\mu_Q)-n(\omega_{Ql+p}+\mu_Q)}
     {i\Omega_m+\omega_{Ql}-\omega_{Ql+p}}\right. \nonumber \\
&&\left.-\frac{n(\omega_{Ql}-\mu_Q)-n(\omega_{Ql+p}-\mu_Q)}
     {i\Omega_m-\omega_{Ql}+\omega_{Ql+p}}
  - \frac{1-n(\omega_{Ql}-\mu_Q)-n(\omega_{Ql+p}+\mu_Q)}
     {i\Omega_m-\omega_{Ql}-\omega_{Ql+p}} \right\},
\end{eqnarray}
where $\exp(-i\Omega_m\beta)=1$ has been used. Substituting Eq. (3.10) into
Eqs. (3.5) and (3.6) and  considering  the  gap equation (2.16), we obtain
the propagator for scalar bound state $\phi^0_S$ in the imaginary-time
formalism
\begin{equation}
\Gamma_{I}^{\phi_S^0}(-i\Omega_m, \stackrel{\rightharpoonup}{p})
=-G/\sum_{Q}g_{QQ}(-\Omega_m^2-{\stackrel{\rightharpoonup}{p}}^2-4m_Q^2)
    K^T_Q(-i\Omega_m, \stackrel{\rightharpoonup}{p}).
\end{equation}
The imaginary-time propagator
$\Gamma_{I}^{\phi_S^0}(-i\Omega_m, \stackrel{\rightharpoonup}{p})$ is defined
at discrete values $-i\Omega_m(m=0, \pm 1, \pm 2, \cdots)$ in the imaginary
axis on the complex energy $p^0$ plane.  The analytic continuation to physical
real values of energy can be made by the replacement [14]
\begin{equation}
-i\Omega_m \longrightarrow p^0 + i\varepsilon p^0, \ \ \ \varepsilon =0_+.
\end{equation}
This means that one can rotate the integral path from the imaginary axis on
the complex $p^0$ plane clockwise to the real axis without meeting any
singularities.  As will be seen later, all  the results derived from Eq. (3.17)
will at least automatically reproduce the expressions of the causal propagators
obtained in usual zero temperature field theory when $T=0$, and this fact
justifies the continuation.  Under the analytic continuation (3.17), we  will
have the substitutions
\begin{equation}
-(\Omega_m^2+{\stackrel{\rightharpoonup}{p}}^2) \longrightarrow
{p^0}^2-{\stackrel{\rightharpoonup}{p}}^2 +i\varepsilon=p^2+i\varepsilon,
\end{equation}
\begin{equation}
A_Q(-i\Omega_m, \stackrel{\rightharpoonup}{p},\stackrel{\rightharpoonup}{l})
\longrightarrow
A_Q(p^0, \stackrel{\rightharpoonup}{p},\stackrel{\rightharpoonup}{l})
=A_Q(p,\stackrel{\rightharpoonup}{l}),
\end{equation}
where
\begin{eqnarray}
A_Q(p,\stackrel{\rightharpoonup}{l})&=&
 \frac{1}{4\omega_{Ql}\omega_{Ql+p}}\left\{
\frac{1-n(\omega_{Ql}+\mu_Q)-n(\omega_{Ql+p}-\mu_Q)}
     {-p^0+\omega_{Ql}+\omega_{Ql+p}-i\varepsilon}
+\frac{n(\omega_{Ql}+\mu_Q)-n(\omega_{Ql+p}+\mu_Q)}
     {-p^0+\omega_{Ql}-\omega_{Ql+p}
            -i\varepsilon \eta(\omega_{Ql}-\omega_{Ql+p})}\right. \nonumber \\
&&\left.-\frac{n(\omega_{Ql}-\mu_Q)-n(\omega_{Ql+p}-\mu_Q)}
     {-p^0-\omega_{Ql}+\omega_{Ql+p}
            +i\varepsilon \eta(\omega_{Ql}-\omega_{Ql+p})}
  - \frac{1-n(\omega_{Ql}-\mu_Q)-n(\omega_{Ql+p}+\mu_Q)}
     {-p^0-\omega_{Ql}-\omega_{Ql+p}+i\varepsilon} \right\},
\end{eqnarray}
with the definition
\begin{equation}
\eta(\omega_{Ql}-\omega_{Ql+p})=\left\{\matrix{
                                1, &{\rm if} \ \ \omega_{Ql}>\omega_{Ql+p},\cr
                               -1, &{\rm if} \ \ \omega_{Ql}<\omega_{Ql+p},\cr}
                               \right.,
\end{equation}
and from  Eq. (3.11),
\begin{equation}
K_Q^T(-i\Omega_m, \stackrel{\rightharpoonup}{p})
\longrightarrow
K_Q^T(p)
=2d_Q(R)\int \frac{d^3l}{(2\pi)^3}A_Q(p,\stackrel{\rightharpoonup}{l}).
\end{equation}
For making a comparison between the results obtained in the imaginary-time and
in the real-time formalism, we may change $A_Q(p,\stackrel{\rightharpoonup}{l})$
into an integral representation. In fact, by the formula
\begin{equation}
\frac{1}{X+i\varepsilon}=\frac{X}{X^2+\varepsilon^2}-i\pi \delta(X)
\end{equation}
and the definition (2.26), we can write
\begin{eqnarray}
A_Q(p,\stackrel{\rightharpoonup}{l})&=&
\int \frac{dl^0}{2\pi}\frac{-i}{[{l^0}^2-\omega^2_{Ql}+i\varepsilon]
                    [{(l^0+p^0)}^2-\omega^2_{Ql+p}+i\varepsilon]} \nonumber \\
&&+\int dl^0\left\{
\frac{\delta({l^0}^2-\omega_{Ql}^2)}{(l^0+p^0)^2-\omega_{Ql+p}^2+i\varepsilon}
              \sin^2\theta(l^0,\mu_Q) +
\frac{\delta[(l^0+p^0)^2-\omega_{Ql+p}^2]}{{l^0}^2-\omega_{Ql}^2+i\varepsilon}
              \sin^2\theta(l^0+p^0,\mu_Q) \right\} \nonumber \\
&&+i2\pi\int dl^0[\theta (l^0)\theta (l^0+p^0)+\theta (-l^0)\theta (-l^0-p^0)]
      \delta({l^0}^2-\omega_{Ql}^2) \delta[(l^0+p^0)^2-\omega_{Ql+p}^2]
      \sin^2\theta(l^0,\mu_Q).
\end{eqnarray}
Applying Eqs. (3.18), (3.22) and (3.24) to Eq. (3.16) and considering the
existence of a more factor $i$ in a four-point function after the analytic continuation  we obtain
the physical propagator for scalar bound state $\phi_S^0$
\begin{eqnarray}
\Gamma_I^{\phi_S^0}(p)&\equiv &i\Gamma_I^{\phi_S^0}
(-i\Omega_m\to p^0+i\varepsilon p^0, \stackrel{\rightharpoonup}{p}) \nonumber \\
&=&-iG/\sum_{Q}g_{QQ}(p^2-4m_Q^2+i\varepsilon)
    K^T_Q(p),
\end{eqnarray}
where
\begin{equation}
K_Q^T(p)=K_Q(p)+H_Q(p)-iS_Q^I(p)
\end{equation}
In Eq. (3.26) the functions
\begin{eqnarray}
K_Q(p)&=&-2d_Q(R)\int\frac{id^4l}{{(2\pi)}^4}\frac{1}{(l^2-m_Q^2+i\varepsilon)
         [(l+p)^2-m_Q^2+i\varepsilon]} \nonumber \\
      &=&\frac{d_Q(R)}{8\pi^2}\int \limits_{0}^{1}dx\left(
          \ln\frac{\Lambda^2+M_Q^2}{M_Q^2}-\frac{\Lambda^2}{\Lambda^2+M_Q^2}
          \right), \ \ M_Q^2=m_Q^2-p^2x(1-x),
\end{eqnarray}
with the four-dimension Euclidean momentum cutoff $\Lambda$,
\begin{equation}
H_Q(p)=4\pi d_Q(R)\int \frac{d^4l}{{(2\pi)}^4}\left\{
         \frac{(l+p)^2-m_Q^2}{{[(l+p)^2-m_Q^2]}^2+\varepsilon^2}+(p\to -p)
         \right\}\delta(l^2-m_Q^2)\sin^2\theta(l^0,\mu_Q),
\end{equation}
and
\begin{eqnarray}
S_Q^I(p)&=&4\pi^2d_Q(R)\int
     \frac{d^4l}{{(2\pi)}^4}\delta(l^2-m_Q^2)\delta[(l+p)^2-m_Q^2] \nonumber \\
   & &  [1-\theta(l^0)\theta(l^0+p^0)-\theta(-l^0)\theta(-l^0-p^0)]
       [\sin^2\theta(l^0,\mu_Q)+\sin^2\theta(l^0+p^0, \mu_Q)].
\end{eqnarray}
When deriving the expression (3.29), we have used the equality
$S^I_Q(p)=S_Q^I(-p)$ coming from the equality $\Gamma_I^{\phi_S^0}(p)=
\Gamma_I^{\phi_S^0}(-p)$. It should be indicated that the expressions (3.26)
-(3.29) for $K_Q^T(p)$ are true for both $\omega_{Ql}>\omega_{Ql+p}$ and
$\omega_{Ql}<\omega_{Ql+p}$ in Eq. (3.20) of
$A_Q(p, \stackrel{\rightharpoonup}{l})$.  By means of the relation
\begin{equation}
g_{QQ}/G=m_Q^2/\sum_{Q}m_Q^2
\end{equation}
derived from Eq. (2.15), we  obtain
\begin{eqnarray}
\Gamma_I^{\phi_S^0}(p)&=&-i\sum_{Q}m_Q^2/
    \sum_{Q}(p^2-4m_Q^2+i\varepsilon)m_Q^2K_Q^T(p)  \nonumber \\
                     &=&-i\sum_{Q}m_Q^2/
    \sum_{Q}(p^2-4m_Q^2+i\varepsilon)m_Q^2[K_Q(p)+H_Q(p)-iS_Q^I(p)].
\end{eqnarray}
It is indicated that the term containing $S_Q^I(p)$ in Eq. (3.31) will make the pole
of $\Gamma^{\phi_S^0}(p)$  possibly become complex. In addition, $K_Q(p)$ in
Eq.(3.27) can also be complex when $p^2>4m_Q^2$.  Denote $K_Q(p)=K_{Qr}(p)-iK_{Qi}$ 
with $K_{Qi}>0$  and let
\[  \sum_{Q}m_Q^2K_{Qr}(p)=k_r,\ \ 
\sum_{Q}m_Q^2K_{Qi}(p)=k_i,\ \ \sum_{Q}m_Q^2H_Q(p)=h,
\ \ \sum_{Q}m_Q^2S_Q^I(p)=s^I,
\]
\begin{equation}\sum_{Q}m_Q^4K_{Qr}(p)=\tilde{k}_r,
\sum_{Q}m_Q^4K_{Qi}(p)=\tilde{k}_i,\ \ \sum_{Q}m_Q^4H_Q(p)=\tilde{h},\ \
\sum_{Q}m_Q^4S_Q^I(p)=\tilde{s}^I.
\end{equation}
We can obtain from Eq. (3.31) the equation to determine the mass of $\phi_S^0$
\begin{equation}
m^2_{\phi_S^0}=p^2=4\frac{\tilde{k}_r+\tilde{h}-i(\tilde{k}_i+\tilde{s}^I)}
                         {k_r+h-i(k_i+s^I)}.
\end{equation}
In the  special case when only single-flavor $Q$ fermions exist (e.g. in
the top-quark condensate scheme [16]) or all the $Q$ fermions are mass
degenerate, we will have the real $m^2_{\phi_S^0}=p^2=4m_Q^2$.  In the other
cases, the solution of $p^2$ (or say $p^0$) will be complex. Denote
$p^0=p^0_r+ip^0_i$, then Eq. (3.33) will become
\begin{equation}
(p^0_r+ip^0_i)^2-{\stackrel{\rightharpoonup}{p}}^2=[a(p)+ib(p)]|_{p^2=
m^2_{\Phi_S^0}}
\end{equation}
with
\[a(p)=4\frac{(k_r+h)(\tilde{k}_r+\tilde{h})+(k_i+s^I)(\tilde{k}_i+\tilde{s}^I)}
             {(k_r+h)^2+(k_i+s^I)^2}, \]
\begin{equation}
b(p)=\frac{(\tilde{k}_r+\tilde{h})(k_i+s^I)-(k_r+h)(\tilde{k}_i+\tilde{s}^I)}
             {(k_r+h)^2+(k_i+s^I)^2}.
\end{equation}
It is easy to find from Eq. (3.35) that $b(p)\equiv 0$ for the cases with both 
single-flavor and mass-degenerate $Q$ fermions and $b(p)$ could be very  small if 
the momentum cutoff $\Lambda$ in the $K_Q(p)$ is very large. Thus we can define the 
squared mass of $\phi_S^0$ by the solution of the real part of Eq. (3.34), i.e.
\begin{equation}
m^2_{\phi_S^0}=p^2_r=a(p_r)=4\left.\frac{(k_r+h)(\tilde{k}_r+\tilde{h})+
               (k_i+s^I)(\tilde{k}_i+\tilde{s}^I)}
             {(k_r+h)^2+(k_i+s^I)^2}\right|_{p=p_r}.
\end{equation}
Since $K_{Qr}(p)$ and $H_Q(p)$ is real and positive [11], and based on the
expression (3.29), the same is true to $S_Q^I(p)$, we can deduce the mass
inequalities from Eq. (3.36)
\begin{equation}
2(m_Q)_{\rm min}\leq m_{\phi_S^0} \leq 2(m_Q)_{\rm max}
\end{equation}
where $(m_Q)_{\rm min}$ and $(m_Q)_{\rm max}$ are respectively the minimal and the 
maximal mass of the $Q$ fermions.  When $0\neq m_U\neq m_D \neq 0$, only the signs 
of inequality are left in Eq. (3.37). In this case,  if we set $m_D=\alpha m_U (\alpha 
>0)$, then the numerator of $b(p)$ in Eq. (3.35) will  become 
$(\alpha^2-\alpha^4)m^6_U[(K_{Ur}+H_U)(K_{Di}+S_D^I) 
-(K_{Dr}+H_D)(K_{Ui}+S_U^I)]$.  Considering the inequalities in Eq. (3.37) and that 
the imaginary parts $K_{Qi}(p)=0$ and $S^I_Q(p)=0$ when $p^2<4m_Q^2$, we can obtain  
that  whether $\alpha <1 (m_D< m_U)$ or $\alpha >1 (m_D> m_U)$  always  have  
$b(p)>0$.  This means that $p^0$ will contain an positive imaginay part
\begin{equation}
p^0_i\simeq \frac{b(p_r)}{2p^0_r},
\end{equation}
hence the amplitude associated with the scalar bound state $\phi_S^0$  will get a 
growth factor $\exp(p^0_i t)$.  In other words, owing to the mass difference
between massive $U$ and $D$ fermions, the scalar bound  state $\phi_S^0$ will encounter 
some effect of fluctuation. It should  be  mentioned that such amplitude growth 
factor of  $\phi_S^0$  always exists in the case with unequal  non-zero masses of 
the two flavors of fermions in a one-generation fermion condensate model even if let 
the temperature $T \to 0$. At finite temperature, it is only  modified by thermal 
effect and displays itself more plainly.  However, such fluctuation effect of 
$\phi_S^0$ is physically  completely negligible considering that $b(p_r)$ will be 
extremely small if the  momentum cut-off $\Lambda$ in the zero temperature loops is 
large enough and that  $\phi_S^0$ generally has a finite decay life in a real model.
\section{Neutral pseudoscalar bound state in the imaginary-time formalism}
The propagator for a neutral pseudoscalar bound state can be calculated by
the four-point amputated Green functions $\Gamma_P^{Q'Q}(-i\Omega_m,
\stackrel{\rightharpoonup}{p})$ for the transition from $(\bar{Q}i\gamma_5Q)$
to $(\bar{Q'}i\gamma_5Q')$.  Based on the pseudoscalar four-fermion coupling
${\cal L}_{4F}^P$ [Eq. (2.4)], in the fermion bubble diagram approximation,
$\Gamma_P^{Q'Q}(-i\Omega_m,\stackrel{\rightharpoonup}{p})$ will obey the
algebraic equations
\begin{equation}
\Gamma_P^{Q'Q''}(-i\Omega_m, \stackrel{\rightharpoonup}{p})[
\delta_{Q''Q}-N_{Q''5}(-i\Omega_m, \stackrel{\rightharpoonup}{p}) g'_{Q''Q}]
=g'_{Q'Q}/2,
\end{equation}
where $2N_{Q5}(-i\Omega_m, \stackrel{\rightharpoonup}{p})$ is the contribution of
the $Q$ fermion loop with pseudoscalar coupling vertices and
\begin{equation}
N_{Q5}(-i\Omega_m, \stackrel{\rightharpoonup}{p})
=-\frac{d_Q(R)}{2}\int \frac{d^3l}{(2\pi)^3}
 T\sum_{n=-\infty}^{\infty}{\rm tr}
 [i\gamma_5S_Q(-i\omega_n+\mu_Q, \stackrel{\rightharpoonup}{l})
  i\gamma_5S_Q(-i\omega_n-i\Omega_m+\mu_Q, \stackrel{\rightharpoonup}{l} +
                                 \stackrel{\rightharpoonup}{p})].
\end{equation}
Similar to Eqs. (3.2), Eqs. (4.1) have the solutions
\begin{equation}
\Gamma_P^{Q'Q}(-i\Omega_m, \stackrel{\rightharpoonup}{p})
=\frac{g'_{Q'Q}}{2{\Delta}'(-i\Omega_m, \stackrel{\rightharpoonup}{p})}
\end{equation}
with
\begin{equation}
{\Delta}'(-i\Omega_m, \stackrel{\rightharpoonup}{p})
=1-\sum_{Q=U,D}g_{QQ}N_{Q5}(-i\Omega_m, \stackrel{\rightharpoonup}{p}).
\end{equation}
The propagator in the imaginary-time formalism for the pseudoscalar bound state
$\phi_P^0$ shown in Eq. (2.9) is
\begin{equation}
\Gamma_{I}^{\phi_P^0}(-i\Omega_m, \stackrel{\rightharpoonup}{p})
=G/2{\Delta}'(-i\Omega_m, \stackrel{\rightharpoonup}{p}).
\end{equation}
By means of Eqs. (2.13), (2.12) and (4.2), we find
\begin{eqnarray}
N_{Q5}(-i\Omega_m, \stackrel{\rightharpoonup}{p})& =& I_Q+
2d_Q(R)\int \frac{d^3l}{(2\pi )^3}T\sum_{n}
\frac{-(\Omega_m^2+{\stackrel{\rightharpoonup}{p}}^2)
      -(\omega_n+i\mu_Q)\Omega_m-\stackrel{\rightharpoonup}{l}\cdot
       \stackrel{\rightharpoonup}{p}}
     {[(\omega_n+i\mu_Q)^2+{\stackrel{\rightharpoonup}{l}}^2+m_Q^2]
      [(\omega_n+\Omega_m+i\mu_Q)^2+{(\stackrel{\rightharpoonup}{l}+
                                      \stackrel{\rightharpoonup}{p})}^2+m_Q^2]}
                                      \nonumber \\
   &=&I_Q-\frac{1}{2}(\Omega_m^2+{\stackrel{\rightharpoonup}{p}}^2)
                K_Q^T(-i\Omega_m, \stackrel{\rightharpoonup}{p}),
\end{eqnarray}
where Eq. (3.9) has been used. Substituting Eq. (4.6) into Eqs. (4.4) and (4.5)
and considering the gap equation (2.16), we obtain
\begin{equation}
\Gamma_{I}^{\phi_P^0}(-i\Omega_m, \stackrel{\rightharpoonup}{p})
=-G/(-\Omega_m^2-{\stackrel{\rightharpoonup}{p}}^2)\sum_{Q}g_{QQ}
    K^T_Q(-i\Omega_m, \stackrel{\rightharpoonup}{p}).
\end{equation}
Then, by the analytic continuation to physical real energy $p^0$ similar to that
made for $\Gamma_{I}^{\phi_S^0}(-i\Omega_m, \stackrel{\rightharpoonup}{p})$, we
obtain the physical propagator for pseudoscalar bound state $\phi_P^0$
\begin{eqnarray}
\Gamma_I^{\phi_P^0}(p)
 &=&-iG/(p^2+i\varepsilon)\sum_{Q}g_{QQ}K^T_Q(p) \nonumber \\
 &=& -i\sum_{Q}m_Q^2/(p^2+i\varepsilon)\sum_{Q}m_Q^2K_Q^T(p),
\end{eqnarray}
where  Eq.(3.30) has been used and $K_Q^T(p)$ is still given by Eqs. (3.26)-(3.29). 
The expression (4.8) indicates that
$\Gamma_I^{\phi_P^0}(p)$ has a single pole at $p^2=0$ thus $\phi_P^0$ is a
massless neutral pseudoscalar Goldstone boson.
In addition, since when $p^2=0$, $H_Q(p)=0$ and $S_Q^I(p)=0$ in $K_Q^T(p)$,
we also have
\begin{equation}
\Gamma_I^{\phi^0_P}(p)=-i\sum_{Q}m_Q^2/(p^2+i\varepsilon)\sum_{Q}m_Q^2K_Q(p),
\ \ \ \     {\rm if} \ \ p^2\rightarrow 0,
\end{equation}
which has the same form as the propagator for $\phi_P^0$ at $T=0$, except that
$m_Q$ now implies the dynamical mass of the $Q$ fermions at temperature $T$
[13]. These results are independent of the mass-difference between the two flavors 
of fermions.
\section{Charged scalar bound states in the imaginary-time formalism}
By Eq. (2.9), the configuration of the charged scalar bound state is
$\phi^-=(\bar{U}\Gamma^-D)$, thus the four-point amputated function for the
transition from $(\bar{U}\Gamma^-D)$ to $(\bar{D}\Gamma^+U)$ will simply
correspond to the propagator  for $\phi^-$ (as well as its hermitian conjugate
$\phi^+$). Based on ${\cal L}_{4F}^C$ in Eq. (2.6), the imaginary-time
propagator $\Gamma_{I}^{\phi^-}(-i\Omega_m, \stackrel{\rightharpoonup}{p})$ for
$\phi^-$  obeys the algebraic equation
\begin{equation}
\Gamma_{I}^{\phi^-}(-i\Omega_m, \stackrel{\rightharpoonup}{p})
=\frac{G}{2}+GL(-i\Omega_m, \stackrel{\rightharpoonup}{p})
\Gamma_{I}^{\phi^-}(-i\Omega_m, \stackrel{\rightharpoonup}{p}),
\end{equation}
from which we find
\begin{equation}
\Gamma_{I}^{\phi^-}(-i\Omega_m, \stackrel{\rightharpoonup}{p})
=G/2[1-GL(-i\Omega_m, \stackrel{\rightharpoonup}{p})],
\end{equation}
where $2L(-i\Omega_m, \stackrel{\rightharpoonup}{p})$ represents the
contribution of the fermion loops bounded by a $U$ fermion and a $D$
fermion propagator with a $\Gamma^+$ and a $\Gamma^-$ coupling vertex, i.e.
we have
\begin{equation}
GL(-i\Omega_m, \stackrel{\rightharpoonup}{p})
=G\frac{d_Q(R)}{2}\int \frac{d^3l}{(2\pi )^3}T\sum_{n}
\frac{-{\rm tr}[\Gamma^-(m_U-\bar{\not l}_U)\Gamma^+(m_D-\bar{\not l}_D
                                                         -\bar{\not\!{p}})]}
     {[(\omega_n+i\mu_U)^2+{\stackrel{\rightharpoonup}{l}}^2+m_U^2]
      [(\omega_n+\Omega_m+i\mu_D)^2+{(\stackrel{\rightharpoonup}{l}+
                                      \stackrel{\rightharpoonup}{p})}^2+m_D^2]}
\end{equation}
with
\begin{equation}
\bar{p}=(\Omega_m, \stackrel{\rightharpoonup}{p}).
\end{equation}
The trace in Eq. (5.3) can be expressed by
\[a=-{\rm tr}[\Gamma^-(m_U-\bar{\not l}_U)\Gamma^+(m_D-\bar{\not l}_D
                                                         -\bar{\not\!{p}})]
   =4\bar{l}_U\cdot (\bar{l}_D+\bar{p})+8\frac{m_U^2m_D^2}{m_U^2+m_D^2},\]
where we have used Eqs. (2.7) and (3.30).  It can be further written by either
\begin{eqnarray}
a\equiv a_U&=&4\left \{(\omega_n+\Omega_m+i\mu_D)^2+{(\stackrel{\rightharpoonup}{l}+
\stackrel{\rightharpoonup}{p})}^2+m_D^2-
[\omega_n+i\mu_U+\Omega_m+i(\mu_D-\mu_U)][\Omega_m+i(\mu_D-\mu_U)] \right.
\nonumber \\
&&\left.  -(\stackrel{\rightharpoonup}{l}+\stackrel{\rightharpoonup}{p})\cdot
\stackrel{\rightharpoonup}{p}+
m^2_D(m_U^2-m_D^2)/(m_U^2+m_D^2)\right \}
\end{eqnarray}
or
\begin{equation}
a\equiv a_D=4\left \{(\omega_n+i\mu_U)^2+{\stackrel{\rightharpoonup}{l}}^2+m_U^2+
(\omega_n+i\mu_U)[\Omega_m+i(\mu_D-\mu_U)]
+\stackrel{\rightharpoonup}{l}\cdot \stackrel{\rightharpoonup}{p}+
m^2_U(m_D^2-m_U^2)/(m_U^2+m_D^2)\right \}.
\end{equation}
From these expressions we obtain
\begin{eqnarray}
GL(-i\Omega_m, \stackrel{\rightharpoonup}{p})
&=&\frac{d_Q(R)}{2}\int \frac{d^3l}{(2\pi )^3}T\sum_{n}
\frac{g_{UU}a_U+g_{DD}a_D}
     {[(\omega_n+i\mu_U)^2+{\stackrel{\rightharpoonup}{l}}^2+m_U^2]
      [(\omega_n+\Omega_m+i\mu_D)^2+{(\stackrel{\rightharpoonup}{l}+
                                      \stackrel{\rightharpoonup}{p})}^2+m_D^2]}
      \\
&=&\sum_{Q=U,D}g_{QQ}I_Q+
 2d_Q(R)\int \frac{d^3l}{(2\pi )^3}T\sum_{n} \nonumber \\
&&  \frac{(g_{DD}-g_{UU})\{(\omega_n+i\mu_U)[\Omega_m+i(\mu_D-\mu_U)]\}+
      \stackrel{\rightharpoonup}{l}\cdot \stackrel{\rightharpoonup}{p} -
      g_{UU}\{[\Omega_m+i(\mu_D-\mu_U)]^2+{\stackrel{\rightharpoonup}{p}}^2\}}
     {[(\omega_n+i\mu_U)^2+{\stackrel{\rightharpoonup}{l}}^2+m_U^2]
      [(\omega_n+\Omega_m+i\mu_D)^2+{(\stackrel{\rightharpoonup}{l}+
                                      \stackrel{\rightharpoonup}{p})}^2+m_D^2]},
\end{eqnarray}
where the result coming from Eq. (3.30)
\begin{equation}
g_{UU}m_D^2-g_{DD}m_U^2=0
\end{equation}
has been used.   Substituting Eq. (5.8) into Eq. (5.2) and considering the gap
equation (2.16),   we obtain
\begin{eqnarray}
\Gamma_{I}^{\phi^-}(-i\Omega_m, \stackrel{\rightharpoonup}{p})
&=&-G/4d_Q(R)\int \frac{d^3l}{(2\pi )^3}\left(
 \{(g_{DD}-g_{UU})i\mu_U(\Omega_m+i\mu_D-i\mu_U+
      \stackrel{\rightharpoonup}{l}\cdot \stackrel{\rightharpoonup}{p}) -
  g_{UU}[(\Omega_m+i\mu_D-i\mu_U)^2+\stackrel{\rightharpoonup}{p}^2] \}
             \right.  \nonumber \\
&& \left.  A_c(-i\Omega_m, \stackrel{\rightharpoonup}{p},
                                 \stackrel{\rightharpoonup}{l})+
    (g_{DD}-g_{UU})(\Omega_m+i\mu_D-i\mu_U)B_c(-i\Omega_m,
      \stackrel{\rightharpoonup}{p},\stackrel{\rightharpoonup}{l})\right),
\end{eqnarray}
where the Matsubara frequency sums
\begin{equation}
A_c(-i\Omega_m, \stackrel{\rightharpoonup}{p},\stackrel{\rightharpoonup}{l})
=T\sum_{n}\frac{1}{[(\omega_n+i\mu_U)^2+\omega^2_{Ul}]
          [(\omega_n+\Omega_m+i\mu_D)^2+\omega^2_{Dl+p}]},
\end{equation}
\begin{equation}
B_c(-i\Omega_m, \stackrel{\rightharpoonup}{p},\stackrel{\rightharpoonup}{l})
=T\sum_{n}\frac{\omega_n}{[(\omega_n+i\mu_U)^2+\omega^2_{Ul}]
          [(\omega_n+\Omega_m+i\mu_D)^2+\omega^2_{Dl+p}]}.
\end{equation}
By  menas of the similar method to calculate
$A_Q(-i\Omega_m, \stackrel{\rightharpoonup}{p},\stackrel{\rightharpoonup}{l})$
in Eq. (3.12), we can express
\begin{equation}
A_c(-i\Omega_m, \stackrel{\rightharpoonup}{p},\stackrel{\rightharpoonup}{l})
=\int_{0}^{\beta}d\tau e^{-i\Omega_m \tau}
         \tilde{\Delta}(\tau, \omega_{Ul}, -\mu_U)
         \tilde{\Delta}(\tau, \omega_{Dl+p}, \mu_D),
\end{equation}
\begin{eqnarray}
B_c(-i\Omega_m, \stackrel{\rightharpoonup}{p},\stackrel{\rightharpoonup}{l})
&=&i\int_{0}^{\beta}d\tau e^{-i\Omega_m \tau}
         \left[\frac{\partial}{\partial \tau}
         \tilde{\Delta}(\tau, \omega_{Ul}, -\mu_U)\right]
         \tilde{\Delta}(\tau, \omega_{Dl+p}, \mu_D) \nonumber \\
&=&-i\mu_U
   A_c(-i\Omega_m, \stackrel{\rightharpoonup}{p},\stackrel{\rightharpoonup}{l})
   +
   D_c(-i\Omega_m, \stackrel{\rightharpoonup}{p},\stackrel{\rightharpoonup}{l})
\end{eqnarray}
with the results
\begin{eqnarray}
\left.\matrix{
A_c(-i\Omega_m, \stackrel{\rightharpoonup}{p},\stackrel{\rightharpoonup}{l}) \cr
D_c(-i\Omega_m, \stackrel{\rightharpoonup}{p},\stackrel{\rightharpoonup}{l}) \cr
}\right\}&=&
\left. \matrix{
-i\omega_{Ul} \cr
1             \cr}\right\}
\frac{1}{4\omega_{Ul}\omega_{Dl+p}}\nonumber \\
&&\left[
\frac{1-n(\omega_{Ul}+\mu_U)-n(\omega_{Dl+p}-\mu_D)}
     {i\Omega_m-(\mu_D-\mu_U)+\omega_{Ul}+\omega_{Dl+p}}+
\frac{n(\omega_{Ul}+\mu_U)-n(\omega_{Dl+p}+\mu_D)}
     {i\Omega_m-(\mu_D-\mu_U)+\omega_{Ul}-\omega_{Dl+p}}\right. \nonumber \\
&&\left.\mp
\frac{n(\omega_{Ul}-\mu_U)-n(\omega_{Dl+p}-\mu_D)}
     {i\Omega_m-(\mu_D-\mu_U)-\omega_{Ul}+\omega_{Dl+p}} \mp
\frac{1-n(\omega_{Ul}-\mu_U)-n(\omega_{Dl+p}+\mu_D)}
     {i\Omega_m-(\mu_D-\mu_U)-\omega_{Ul}-\omega_{Dl+p}}
\right].
\end{eqnarray}
In the derivation of Eq. (5.14) we have  used the antiperiodicity condition
(2.21) of $\tilde{\Delta}(\tau, \omega_{Ql}, \mu_Q)$. \\
For analytic continuation of the definition region of the propagator for
$\phi^-$  from $-i\Omega_m$ to the real axis of $p^0$, we will make the
replacement
\begin{equation}
-i\Omega_m +\mu_D-\mu_U\longrightarrow p^0 +i\varepsilon p^0, \ \ \
                     \varepsilon =0_+,
\end{equation}
considering that the charged scalar bound state $\phi^-=(\bar{U}\Gamma^-D)$
composed of the $U$ antifermions and  the $D$ fermions may have the chemical
potential $\mu_D-\mu_U$. Correspondingly, we will have the following
substitutions:
\begin{equation}
A_c(-i\Omega_m, \stackrel{\rightharpoonup}{p},\stackrel{\rightharpoonup}{l})
\rightarrow  A_c(p,\stackrel{\rightharpoonup}{l}), \ \ \
D_c(-i\Omega_m, \stackrel{\rightharpoonup}{p},\stackrel{\rightharpoonup}{l})
\rightarrow  D_c(p,\stackrel{\rightharpoonup}{l}).
\end{equation}
For the purpose of making a comparison between the results in the
imaginary-time formalism and in the real-time formalism, we will express
$A_c(p,\stackrel{\rightharpoonup}{l})$ and
$D_c(p,\stackrel{\rightharpoonup}{l})$ respectively by an integral over the real
$l^0$. In this way, it is found that
\begin{equation}
A_c(p,\stackrel{\rightharpoonup}{l})=
\int \frac{dl^0}{2\pi}\tilde{A}_c(p,l^0,\stackrel{\rightharpoonup}{l}),
\end{equation}
\begin{equation}
D_c(p,\stackrel{\rightharpoonup}{l})=
\int \frac{dl^0}{2\pi}il^0 \tilde{A}_c(p,l^0,\stackrel{\rightharpoonup}{l}),
\end{equation}
where
\begin{eqnarray}
\tilde{A}_c(p,l^0,\stackrel{\rightharpoonup}{l})
&=&\frac{-i}{(l^2-m_U^2+i\varepsilon)
         [(l+p)^2-m_D^2+i\varepsilon]} \nonumber \\
&&+2\pi \frac{\delta(l^2-m_U^2)}{(l+p)^2-m_D^2+i\varepsilon}
                    \sin^2\theta(l^0,\mu_U) +
   2\pi \frac{\delta[(l+p)^2-m_D^2]}{l^2-m_U^2+i\varepsilon}
                    \sin^2\theta(l^0+p^0,\mu_D) \nonumber \\
&&+i4\pi^2 \delta(l^2-m_U^2)\delta[(l+p)^2-m_D^2]
       [\theta(l^0)\theta(l^0+p^0)+\theta(-l^0)\theta(-l^0-p^0)]\nonumber \\
&&  [\theta(\omega_{Ul}-\omega_{Dl+p})\sin^2\theta(l^0,\mu_U)+
          \theta(\omega_{Dl+p}-\omega_{Ul})\sin^2\theta(l^0+p^0, \mu_D)].
\end{eqnarray}
By means of Eqs (5.10), (5.14) and (5.16)-(5.19), we can analytically continue
the imaginary-time propagator for $\phi^-$  to physical propagator, i.e.
\begin{equation}
i\Gamma_I^{\phi^-}(-i\Omega_m, \stackrel{\rightharpoonup}{p})
\rightarrow  \Gamma^{\phi^-}(p)
\end{equation}
with
\begin{eqnarray}
\Gamma_I^{\phi^-}(p)
&=&-iG/4d_Q(R)\int \frac{d^3l}{(2\pi)^3}\{
   [(g_{DD}-g_{UU})
   \stackrel{\rightharpoonup}{l}\cdot \stackrel{\rightharpoonup}{p}+
   g_{UU}(p^2+i\varepsilon)]
   A_c(p,\stackrel{\rightharpoonup}{l}) +
   (g_{DD}-g_{UU})ip^0 D_c(p,\stackrel{\rightharpoonup}{l}) \} \nonumber \\
&=&-iG/4d_Q(R)\int \frac{d^4l}{(2\pi)^4}
      [(g_{UU}-g_{DD}) l\cdot p+ g_{UU}(p^2+i\varepsilon)]
      \tilde{A}_c(p, l^0, \stackrel{\rightharpoonup}{l}) \nonumber \\
&=&-i\sum_{Q}m_Q^2/4d_Q(R)\int \frac{d^4l}{(2\pi)^4}
      [(m_U^2-m_D^2) l\cdot p+ m_U^2 (p^2+i\varepsilon)]
      \tilde{A}_c(p, l^0, \stackrel{\rightharpoonup}{l}),
\end{eqnarray}
where Eq. (3.30) has been used once again. Substituting Eq. (5.20) into
Eq. (5.22) and using the formula (3.23), we finally obtain the physical
propagator  for the charged scalar bound state $\phi^-$
\begin{equation}
\Gamma_I^{\phi^-}(p)=-i/ \left\{
(p^2+i\varepsilon)\left[K_{UD}(p)+H_{UD}(p)\right]+
E_{UD}(p)-i(p^2-\bar{M}^2) S^I_{UD}(p)\right\},
\end{equation}
where
\begin{equation}
\bar{M}^2={(m_U^2-m_D^2)}^2/(m_U^2+m_D^2),
\end{equation}
\begin{eqnarray}
K_{UD}(p)&=&\frac{d_Q(R)}{4\pi^2}\int \limits_{0}^{1}dx
        \frac{m_U^2(1-x)+m_D^2x}{m_U^2+m_D^2}\left[
        \ln \frac{\Lambda^2+M_{UD}^2(p)}{M_{UD}^2(p)}-
            \frac{\Lambda^2}{\Lambda^2+M_{UD}^2(p)}\right],\nonumber \\
         && M_{UD}^2(p)=m_U^2(1-x)+m_D^2x-p^2x(1-x),
\end{eqnarray}
\begin{equation}
H_{UD}(p)=4\pi d_Q(R)\int \frac{d^4l}{{(2\pi)}^4}\left\{
\frac{(l+p)^2-m_D^2}{[(l+p)^2-m_D^2]^2+\varepsilon^2}
\delta(l^2-m_U^2)\sin^2\theta(l^0, \mu_U)+
(p\rightarrow -p, m_U\leftrightarrow m_D, \mu_U\leftrightarrow \mu_D)
\right\},
\end{equation}
\begin{eqnarray}
E_{UD}(p)&=&4\pi d_Q(R)\frac{m_U^2-m_D^2}{m_U^2+m_D^2}
\int \frac{d^4l}{{(2\pi)}^4} \nonumber \\
&&\left\{
\frac{[(l+p)^2-m_U^2][(l+p)^2-m_D^2]}{[(l+p)^2-m_D^2]^2+\varepsilon^2}
\delta(l^2-m_U^2)\sin^2\theta(l^0, \mu_U)
-(p\rightarrow -p, m_U\leftrightarrow m_D, \mu_U\leftrightarrow \mu_D)
\right\},
\end{eqnarray}
\begin{eqnarray}
S^I_{UD}(p)&=&4\pi^2 d_Q(R)\int \frac{d^4l}{{(2\pi)}^4}
          \delta(l^2-m_U^2)\delta[(l+p)^2-m_D^2] \nonumber \\
   &&\left \{ \sin^2\theta(l^0, \mu_U)+\sin^2\theta(l^0+p^0, \mu_D)
       -2[\theta(l^0)\theta(l^0+p^0)+\theta(-l^0)\theta(-l^0-p^0)]\right.
       \nonumber \\
   &&\left.[\theta(\omega_{Ul}-\omega_{Dl+p})\sin^2\theta(l^0, \mu_U)+
      \theta(\omega_{Dl+p}-\omega_{Ul})\sin^2\theta(l^0+p^0, \mu_D)]\right \}.
\end{eqnarray}
As for the key question under what condition
$\phi^{\mp}$ could be massless bound states, we can answer it in two cases
\\
\indent (1) $m_U=m_D=m_Q$. For the mass-degenerate $U$ and $D$ fermions, we
will have $K_{UD}(p)=K_Q(p)$, $H_{UD}(p)=H_Q(p)$,
$S^I_{UD}(p)=S^I_Q(p)$, $E_{UD}(p)=0$, and $\bar{M}^2=0$, thus
\begin{equation}
\Gamma_I^{\phi^-}(p)=-i/(p^2+i\varepsilon)\left[K_Q(p)+H_Q(p)-iS^I_Q(p)\right].
\end{equation}
Eq. (5.29) implies that $p^2=0$ is the single pole of $\Gamma^{\phi^-}(p)$
hence $\phi^-$ and $\phi^+$ are both massless scalar bound states and can be
identified with the charged Nambu-Goldstone bosons. As was indicated in
Sec. III, when $p\to 0$, no pinch singularity could emerge from $S^I_Q(p)$.\\
\indent (2) $m_U\neq m_D$.  For the mass-nondegenerate $U$ and $D$ fermions,
we will have $E_{UD}\neq 0$ and $\bar{M}^2\neq 0$, thus the pole of
$\Gamma_I^{\phi^-}(p)$ will be determined by the equation
\begin{equation}
p^2= - \frac{E_{UD}(p)+i\bar{M}^2S^I_{UD}(p)}
            {K_{UD}(p)+H_{UD}(p)-iS^I_{UD}(p)}.
\end{equation}
Hence, the masses of $\phi^-$ and $\phi^+$ at finite temperature are not equal to 
zeros.  However, as long as the momentum cutoff $\Lambda$ in $K_{UD}(p)$ is large 
enough, the single pole of $\Gamma_I^{\phi^-}(p)$ could still be approximately at 
$p^2=0$. On the other hand, when $p^2=0$ and 
$p^0=|\stackrel{\rightharpoonup}{p}|\rightarrow 0$, both
$E_{UD}(p)$ and $S^I_{UD}(p)$ in the numerator of the right-handed side of
Eq. (5.30) approach to zeros.  Therefore, at low energy scales it is still
possible that $\phi^-$ and $\phi^+$  are considered as approximate massless
bound states and identified with the charged Nambu-Goldstone bosons.
\section{Physical propagators for scalar bound states in the real-time formalism}
The propagators for scalar bound states in the real-time formalism were discussed 
in Ref. [11].  However, the propagators are defined there by 
$\Gamma_B^{Q'1Q1} (p)(B=S,P)$ and $\Gamma_{\phi^-}^{11}(p)$ respectively for
neutral scalar, pseudoscalar and charged scalar bound states which are the four-
point amputated functions for the transition between the physical field 
configurations e.g. $(\bar{Q}Q)^{(1)}$ and ${(\bar{Q'}Q')}^{(1)}$ etc..
These definitions can give the main physical features of the bound states including 
their masses but also lead some extra result, e.g. a neutral scalar bound state 
$\phi_S^0$ could have double masses due thermal fluctuation. In addition, the
resulting expressions of the propagators for $\phi_S^0$,$\phi_P^0$ and $\phi^{\mp}$
have quite explicit difference with Eqs. (3.31), (4.8) and (5.23) obtained here in 
the imaginary-time formalism. For seeking a closer correspondence between the
physical propagators for the above scalar bound states in the real-time and the 
imaginary-time formalism, following Ref.[12], we will redefine the physical 
propagators for $\phi_S^0$, $\phi_P^0$ and $\phi^{\mp}$ by diagonalization of 
corresponding matrix propagators.  These matrix propagators have in fact been 
obtained in Ref. [11].
  
First let us deal with the case of the neutral scalar bound state.   By Eqs. (3.3) 
and (3.7) in Ref.[11], the matrix of the four-point amputated functions for the 
transition from $(\bar{Q}Q)^{(a)}$ and ${(\bar{Q'}Q')}^{(b)}$ can be
expressed by
\begin{equation}
\Gamma_S^{Q'bQa}(p)=\frac{g_{Q'Q}}{G}\Gamma^{\phi_S^0ba}(p), \; \; Q',Q=U,D,\;\;b,a=1,2
\end{equation}
where $\Gamma^{\phi_S^0ba}(p)\; (b,a=1,2)$ is the matrix propagator for the neutral 
scalar bound state $\phi_S^0$ given in Eq. (2.9), its explicit form is
\begin{eqnarray}
\left(\matrix{\Gamma^{\phi_S^011}(p)&\Gamma^{\phi_S^012}(p)\cr \Gamma^{\phi_S^021}(p)& \Gamma^{\phi_S^022}(p) \cr}\right)&=&
\frac{-i\sum_{Q}m_Q^2}{[p^2(k+h)-4(\tilde{k}+\tilde{h})]^2+(p^2s-4\tilde{s})^2
-(p^2r-4\tilde{r})^2}  \nonumber \\
&&\left(\matrix{(p^2-i\varepsilon)(k+h+is)-4(\tilde{k}+\tilde{h}+i\tilde{s})
   &-i(p^2r-4\tilde{r})\cr
   -i(p^2r-4\tilde{r})&
     -(p^2+i\varepsilon)(k+h-is)+4(\tilde{k}+\tilde{h}-i\tilde{s})\cr}\right)
\end{eqnarray}
In Eq. (6.2) the denotations $k$, $h$, $\tilde{k}$ and $\tilde{h}$ have been given 
by Eq.(3.32) and
\begin{equation} 
s=\sum_{Q}m_Q^2S_Q(p),\; r=\sum_{Q}m_Q^2R_Q(p),\;
\tilde{s}=\sum_{Q}m_Q^4S_Q(p),\; \tilde{r}=\sum_{Q}m_Q^4R_Q(p)
\end{equation}
where $K_Q(p)$ and $H_Q(p)$ are expressed by Eqs. (3.27) and (3.28) and
\begin{equation}
S_Q(p)=4\pi^2d_Q(R)\int
       \frac{d^4l}{{(2\pi)}^4}\delta(l^2-m_Q^2)\delta[(l+p)^2-m_Q^2]
       [\sin^2\theta(l^0+p^0,\mu_Q)\cos^2\theta(l^0,\mu_Q)+\cos^2\theta(l^0+p^0,
       \mu_Q)\sin^2\theta(l^0, \mu_Q)],
\end{equation}
and
\begin{equation}
R_Q(p)=2\pi^2d_Q(R)\int
\frac{d^4l}{{(2\pi)}^4}\delta(l^2-m_Q^2)\delta[(l+p)^2-m_Q^2]
\sin 2\theta(l^0,\mu_Q)\sin 2\theta(l^0+p^0, \mu_Q).
\end{equation}
with $\sin^2\theta(l^0, \mu_Q)$ given by Eq. (2.26). In deriving Eq. (6.2), the gap 
equation (2.27) in the real-time formalism has been used. The matrix 
$\Gamma^{\phi_S^0 ba}(p) (b,a=1,2)$ can be diagonalized by a thermal matrix $M_S$,
i.e.
\begin{equation}
\left(\matrix{\Gamma^{\phi_S^011}(p)&\Gamma^{\phi_S^012}(p)\cr \Gamma^{\phi_S^021}(p)& \Gamma^{\phi_S^022}(p) \cr}\right)=M_S^{-1}
\left(\matrix{\Gamma_R^{\phi_S^0}(p)&0\cr 0& \Gamma_R^{\phi_S^0*}(p) \cr}\right)
M_S^{-1}
\end{equation}
where 
\begin{equation}
M_S=\left(\matrix{\cosh \theta_S & \sinh \theta_S\cr
                  \sinh \theta_S &  \cosh \theta_S\cr}\right)
\end{equation}
with
\begin{equation}
\cosh\theta_S=\frac{1}{\sqrt{2}}\left(\frac{S}{\sqrt{S^2-
R^2}}+1\right)^{1/2},\;\;
\sinh\theta_S=\frac{1}{\sqrt{2}}\left(\frac{S}{\sqrt{S^2-R^2}}-1\right)^{1/2}
\end{equation}
and 
\begin{equation}
S=p^2s-4\tilde{s},\; R=p^2r-4\tilde{r}
\end{equation}
We indicate that since $S_Q(p)\pm R_Q(p)\geq 0$ [11] and $ S_Q(p)= R_Q(p)= 0$ when 
$p^2<4m_Q^2$, it can be deduced that $S^2-R^2=\sum_{Q}m_Q^2(p^2-4m_Q^2)(S_Q+R_Q)
\sum_{Q'}m_{Q'}^2(p^2-4m_{Q'}^2)(S_{Q'}-R_{Q'})\geq 0$ no matter what value $p^2$ 
could take.
$\Gamma_R^{\phi_S^0}(p)$ in Eq. (6.6) will be  identified with the physical 
propagator for $\phi_S^0$ in the real-time formalism and has the expression
\begin{equation}
\Gamma_R^{\phi_S^0}(p)=-i\sum_{Q}m_Q^2/\{[(p^2+i\varepsilon)[k+h-is\sqrt{1-
R^2/S^2}]-4[\tilde{k}+\tilde{h}-i\tilde{s}\sqrt{1-R^2/S^2}]\}
\end{equation}
On the other hand, by means of Eq. (3.32),  $\Gamma_I^{\phi_S^0}(p)$ in Eq. (3.31) 
can
be written as
\begin{equation}
\Gamma_I^{\phi_S^0}(p)=-i\sum_{Q}m_Q^2/[(p^2+i\varepsilon)(k+h-is^I)-
4(\tilde{k}+\tilde{h}-i\tilde{s}^I)]
\end{equation}
Comparing Eq. (6.10) with Eq. (6.11), we find that the physical propagators for 
$\phi_S^0$ in the two formalisms now have quite similar form, except that the 
replacements $s\sqrt{1-R^2/S^2}\to s^I$ and $\tilde{s}\sqrt{1-R^2/S^2} \to 
\tilde{s}^I$ in the imaginary parts of their denominators must be made when transiting
from the real-time formalism to the imaginary-time formalism.  It has been known 
for some time that for an amputated Green function, the results calculated in the
two formalisms of thermal field theory have generally some difference [17].  In 
present case, we note that the difference appears only in imaginary parts of the
denominators of the propagators which does not affects the main physical conclusions
from the propagators.  This means that all the conclusions of the mass of $\phi_S^0$
 deduced from $\Gamma_I^{\phi_S^0}(p)$ in Sec. 3 remain to be valid for 
$\Gamma_R^{\phi_S^0}(p)$, including that the possible  fluctuation effect of the
amplitude of $\phi_S^0$ is also kept qualitatively.  In particular, by redefinding 
the physical propagator  $\Gamma_R^{\phi_S^0}(p)$ for $\phi_S^0$ by diagonalization 
of corresponding matrix propagator, we no longer meet the problems of doubling of 
$m_{\phi_S^0}$ originated from the definition of the propagator for $\phi_S^0$ by
$\Gamma^{\phi_S^011}(p)$.  We also point out that the imaginay parts of the 
denominators of both $\Gamma_R^{\phi_S^0}(p)$ and $\Gamma_I^{\phi_S^0}(p)$ may be 
identical and equal to zeros if $0\leq p^2<4(m_Q)_{\rm min}^2$, where $(m_Q)_{\rm 
min}$ is the smallest mass of the $Q$-fermions. In that case,  we have real $k$ and 
$\tilde{k}$, and  $s=\tilde{s}=r=\tilde{r}=0$ thus $S=R=0$ and $s^I=\tilde{s}^I=0$
because the $\delta$-function product factor $\delta(l^2-m_Q^2)\delta[(l+p)^2-
m_Q^2]$ in the integrand of each $S_Q(p)$ and $R_Q(p)$ will become zero if $0\leq 
p^2< 4m_Q^2$. In the other case the imaginary parts of the denominators of  
$\Gamma_R^{\phi_S^0}(p)$ and $\Gamma_I^{\phi_S^0}(p)$  will generally different. 
We note that , different from the case of the gap equation, the appearance of such 
difference here is related to the fact that  we are dealing with the four-point 
amputated functions which are determined by the loops bounded by two fermion 
propagators.   The difference  could be technically attributed to
different order of analytic continuation of a Green function for discrete Matsubara 
frequencies in the two formalisms [12]. But it is more possible [18] that 
appearance of the difference is due to that we are calculating different functions
e.g. $\Gamma_I^{\phi_S^0}(p)$ and $\Gamma_R^{\phi_S^0}(p)$ in the two formalisms.
However, it should be pointed out that at present the two formalisms of thermal field 
theory are not used {\it as usual}, by the terminology in Ref.[18].  First, in the 
imaginary-time formalism, owing to the analytic continuation (3.17) used, 
$\Gamma_I^{\phi_S^0}(p)$ is now a time-ordered  i.e. physical propagator, not a 
retarded one. Thus it is not difficult to understand why $\Gamma_I^{\phi_S^0}(p)$ 
and $\Gamma_R^{\phi_S^0}(p)$ may have almost identical expressions except the 
imaginary parts in their denominators. Next, we note that in the fermion bubble 
approximation, the amputated four-point functions  induced by the four-fermion 
interactions are essentially determined by one fermion loops, and the calculations 
of them may formally be reduced  the ones of usual two-point functions, but the former 
is rather different from the latter. This is because we are dealing with problem of 
the scalar bound states whose propagators correspond to only pure amputated functions 
without external particle legs. In this case, the thermal matrix $M_S$ [Eqs. (6.7) 
and (6.8)] which diagonalizes the amputated  four-point functions matrix (6.6) in 
the real-time formalism is different from the thermal matrix of a free scalar particle 
and can only be considered as an "effective" one in the case of bound states.  
Obviously, some more work is needed before any definite relation between 
$\Gamma_I^{\phi_S^0}(p)$ and $\Gamma_R^{\phi_S^0}(p)$ is found, for instance, by 
some more formal calculations. At present, it seems to be premature for us be able 
to make a judgement on that in which circumstances, which  proper function should 
be considered.  Hence an alternative and more realistic way to treat such difference 
between the two formalisms is to compare the results obtained by respectively using 
the propagators in the two formalisms in the  same physical problem and examine 
whether the difference indeed give some physically unequal description or not. 

Next we turn to the case of pseudoscalar bound state mode and make  similar discussions.  By  means of Eqs. (4.3) and (4.4), the matrix of the four-point amputated functions for the transition from $(\bar{Q}i\gamma_5Q)^{(a)}$ to $(\bar{Q'}i\gamma_5 Q')^{(b)}$ can be expressed by [11]
\begin{equation}
\Gamma_P^{Q'bQa}(p)=\frac{{g'}_{Q'Q}}{G}\Gamma^{\phi_P^0ba}(p), \; \; Q',Q=U,D,\;\;b,a=1,2
\end{equation}
where $\Gamma^{\phi_P^0ba}(p)\; (b,a=1,2)$ is the matrix propagator for the neutral pseudoscalar bound state $\phi_P^0$ in Eq. (2.9) and has the following explicit expression
\begin{equation}
\left(\matrix{\Gamma^{\phi_P^011}(p)&\Gamma^{\phi_P^012}(p)\cr
 \Gamma^{\phi_P^021}(p)& \Gamma^{\phi_P^022}(p) \cr}\right)=
\frac{-i\sum_{Q}m_Q^2}{[(p^2)^2+\varepsilon^2][(k+h)^2+s^2-r^2]}
\left(\matrix{(p^2-i\varepsilon)(k+h+is)
   &-ip^2r\cr
   -ip^2r&
     -(p^2+i\varepsilon)(k+h-is)\cr}\right)
\end{equation}
The matrix $\Gamma^{\phi_P^0ba}(p)\; (b,a=1,2)$ can be diagonalized by a thermal matrix $M_P$, i.e.
\begin{equation}
\left(\matrix{\Gamma^{\phi_P^011}(p)&\Gamma^{\phi_P^012}(p)\cr \Gamma^{\phi_P^021}(p)& \Gamma^{\phi_P^022}(p) \cr}\right)=M_P^{-1}
\left(\matrix{\Gamma_R^{\phi_P^0}(p)&0\cr 0& \Gamma_R^{\phi_P^0*}(p) \cr}\right)
M_P^{-1}
\end{equation}
where
\begin{equation}
M_P=\left(\matrix{\cosh \theta_P & \sinh \theta_P\cr
                  \sinh \theta_P &  \cosh \theta_P\cr}\right)
\end{equation}
with
\begin{equation}
\cosh\theta_P=\frac{1}{\sqrt{2}}\left(\frac{s}{\sqrt{s^2-r^2}}+1\right)^{1/2},\;\;
\sinh\theta_s=\frac{1}{\sqrt{2}}\left(\frac{s}{\sqrt{s^2-r^2}}-1\right)^{1/2}
\end{equation}
Here we again have $s^2-r^2\geq 0$ due to $S_Q(p)\pm R_Q(p) \geq 0$ [11]. The physical 
propagator for $\phi_P^0$ in the real-time formalism will become 
\begin{equation}
\Gamma_R^{\phi_P^0}(p)=-i\sum_{Q}m_Q^2/(p^2+i\varepsilon)[k+h-is\sqrt{1-r^2/s^2}]
\end{equation}
It can be compared with  $\Gamma_I^{\phi_P^0}(p)$, the propagator  for $\phi_P^0$ 
in the imaginary-time formalism which, by Eqs. (4.8), (3.26) and (3.23), will have 
the expression
\begin{equation}
\Gamma_I^{\phi_P^0}(p)=-i\sum_{Q}m_Q^2/(p^2+i\varepsilon)(k+h-is^I)
\end{equation}
We see that $\Gamma_R^{\phi_P^0}(p)$ and $\Gamma_I^{\phi_P^0}(p)$ have very similar 
expression except that in the imaginary parts of their denominators the replacement
$s\sqrt{1-r^2/s^2}\to s^I$ must be made when transiting from the real-time formalism 
to the imaginary-time formalism. Such difference does not change the main physical 
conclusion reached from $\Gamma_R^{\phi_P^0}(p)$ or $\Gamma_I^{\phi_P^0}(p)$: 
$\phi_P^0$ is a massless neutral pseudoscalar particle and can be identified with
a Nambu-Goldstone boson of electroweak symmetry breaking. We also note that when 
$0\leq p^2<4(m_Q)_{\rm min}^2$, $\Gamma_R^{\phi_P^0}(p)$ and
$\Gamma_I^{\phi_P^0}(p)$ will be real and idetical since $s=r=s^I=0$ in this case.
In particular, when $p^2\to 0$, both  have the same form as the propagator for 
$\phi_P^0$ at $T=0$.

Lastly we apply the parallel discussions to the charged scalar bound state mode.
By Eqs. (5.3) and (5.4) in Ref. [11], the matrix of the four-point amputated functions 
for the transition from  $(\bar{U}\Gamma^-D)^{(a)}$ to $(\bar{D}\Gamma^+U)^{(b)}$
is just the matrix propagator $\Gamma_{\phi^-}^{ba}\; (b,a=1,2)$ for $\phi^-
=(\bar{U}\Gamma^-D)$ and its elements can be expressed as follows:
\[
\Gamma_{\phi^-}^{11}(p)=-i\frac
{(p^2-i\varepsilon)[K_{UD}(p)+H_{UD}(p)+iS_{UD}(p)]+E_{UD}(p)-i\bar{M}^2S_{UD}(p)}
{\{p^2[K_{UD}(p)+H_{UD}(p)]+E_{UD}(p)\}^2+(p^2-\bar{M}^2)^2[S_{UD}^2(p)-R_{UD}^2(p)]}= [\Gamma_{\phi^-}^{22}(p)]^*\]
\begin{equation}
\Gamma_{\phi^-}^{12}(p)=\frac
{- e^{\beta(\mu_U-\mu_D)/2}\,(p^2-\bar{M}^2)R_{UD}(p)}
{\{p^2[K_{UD}(p)+H_{UD}(p)]+E_{UD}(p)\}^2+(p^2-\bar{M}^2)^2[S_{UD}^2(p)-R_{UD}^2(p)]}= e^{\beta(\mu_U-\mu_D)}\, \Gamma_{\phi^-}^{21}(p)
\end{equation}
where $\bar{M}^2$, $K_{UD}(p)$, $H_{UD}(p)$ and $E_{UD}(p)$ have been given by Eqs.
(5.24)-(5.27) respectively, and by Eqs. (5.11) and (5.12) in Ref. [11], we have
\begin{eqnarray}
S_{UD}(p)&=&4\pi^2 d_Q(R)\int \frac{d^4l}{{(2\pi)}^4}
          \delta(l^2-m_U^2)\delta[(l+p)^2-m_D^2] \nonumber \\
          && \left\{\sin^2\theta(l^0, \mu_U)\cos^2\theta(l^0+p^0, \mu_D)+
          \cos^2\theta(l^0, \mu_U)\sin^2\theta(l^0+p^0, \mu_D)
          \right\},
\end{eqnarray}
\begin{equation}
R_{UD}(p)=2\pi^2 d_Q(R)\int \frac{d^4l}{{(2\pi)}^4}
          \delta(l^2-m_U^2)\delta[(l+p)^2-m_D^2]
          \sin 2\theta(l^0, \mu_U)\sin 2\theta(l^0+p^0, \mu_D).
\end{equation}
By means of a thermal matrix $M_C$, the matrix $\Gamma_{\phi^-}^{ba}\; (b,a=1,2)$
 can be diagonalized, i.e.
\begin{equation}
\left(\matrix{\Gamma_{\phi^-}^{11}&\Gamma_{\phi^-}^{12}\cr
\Gamma_{\phi^-}^{21}&\Gamma_{\phi^-}^{22}\cr}\right)=M_C^{-1}\left(\matrix{
\Gamma_R^{\phi^-}(p)&0\cr
 0&    {\Gamma_R^{\phi^-}}^*(p)\cr}\right)M_C^{-1}
\end{equation} 
where 
\begin{equation}
M_C=\left(\matrix{\cosh \theta_C & e^{\alpha}\sinh \theta_C\cr
                  e^{-\alpha}\sinh \theta_C &  \cosh \theta_C\cr}\right)
\end{equation}
with
\begin{equation}
\alpha=\beta(\mu_U-\mu_D)/2, \;\;
\cosh\theta_C=\frac{1}{\sqrt{2}}\left[\frac{S_{UD}(p)}{\sqrt{S_{UD}^2(p)-
R_{UD}^2(p)}}+1\right]^{1/2},\;\;
\sinh\theta_C=\frac{1}{\sqrt{2}}\left[\frac{S_{UD}}{\sqrt{S_{UD}^2(p)-
R_{UD}^2(p)}}-1\right]^{1/2}
\end{equation}
It is easy from Eqs. (6.20) and (6.21) to verify that 
\begin{equation}
S_{UD}(p)\pm R_{UD}(p)=4\pi^2d_Q(R)\int \frac{d^4l}{{(2\pi)}^4}
                \delta(l^2-m_U^2)\delta[{(l+p)}^2-m_D^2]
                \sin^2[\theta(l^0,\mu_U)\pm \theta(l^0+p^0, \mu_D)]\geq 0.
\end{equation}
thus we must have $S_{UD}^2(p)- R_{UD}^2(p)\geq 0$.  $\Gamma_R^{\phi^-}(p)$ in 
Eq. (6.22) is now defined as the physical propagator for $\phi^-$ in the real-time 
formalism and can be expressed by
\begin{equation}
\Gamma_R^{\phi^-}(p)=-i/\left\{(p^2+i\varepsilon)[K_{UD}(p)+H_{UD}(p)]+E_{UD}(p)-i(p^2-
\bar{M}^2)\sqrt{S_{UD}^2(p)-R_{UD}^2(p)}\right\}
\end{equation}
It has very similar form to $\Gamma_I^{\phi^-}(p)$ in Eq. (5.23), the propagator for 
$\phi^-$ in the imaginary-time formalism, but for the transition 
$\Gamma_R^{\phi^-}(p) \to \Gamma_I^{\phi^-}(p)$ we must make the replacement
$\sqrt{S_{UD}^2(p)-R_{UD}^2(p)} \to S^I_{UD}(p)$ in the imaginary parts of their
denominators.  Obviously, all the conclusions coming from $\Gamma_R^{\phi^-}(p)$ 
will be the same as those from $\Gamma_I^{\phi^-}(p)$ in Sec.V. When $m_U=m_D=m_Q$
and $\mu_U=\mu_D=\mu_Q$, we will have $E_{UD}=\bar{M}^2=0$, $K_{UD}=K_Q$,
$H_{UD}=H_Q$, $S_{UD}=S_Q$, and $R_{UD}=R_Q$, thus we get
\begin{equation}
\Gamma_R^{\phi^-}(p)=-i/(p^2+i\varepsilon)\left[K_Q(p)+H_Q(p)-i\sqrt{S_Q^2(p)-
R_Q^2(p)}\right]
\end{equation}
which has identical form to $\Gamma_R^{\phi_P^0}(p)$ in this case. The
thermal matrix element
\[\cosh\theta_C=\frac{1}{\sqrt{2}}\left[S_Q(p)/\sqrt{S_Q^2(p)-
R_Q^2(p)}+1\right]^{1/2}\]
will also coincide with $\cosh \theta_P$.
In addition, it can be proven that
\begin{equation}
\delta(l^2-m_U^2)\delta[{(l+p)}^2-m_D^2]=0, \; \; {\rm when}\;
 (m_U-m_D)^2\leq   p^2<(m_U+m_D)^2,
\end{equation}
thus $S_{UD}(p)$, $R_{UD}(p)$ and $S_{UD}^I(p)$ are all equal to zeroes and
these will make $ \Gamma_R^{\phi^-}(p)$ and  $\Gamma_I^{\phi^-}(p)$ become
identical in this case.
\section{Conclusions}
In the imaginary-time formalism of thermal field theory we have reexamined the
Nambu-Goldstone mechanism of electroweak symmetry breaking at finite
temperature in a one-generation fermion condensate scheme and compared the
results with those obtained in the real-time formalism through some redefined 
physical propagators for scalar bound states.  By means of the
Schwinger-Dyson equation in the fermion bubble diagram approximation, it is
obtained that the propagators for scalar bound states have very similar forms in the  
two formalisms, except that  the imaginary parts of the denominators of the 
propagators have some differences when the momenta squared of the bound states are 
within some given ranges.  However, these differences do not change the common 
essential conclusions reached in the two formalisms. These conclusions are as 
follows.\\
\indent 1) When the two flavors of the one generation of fermions are mass-degenerate,
  at the temperature $T$ below the symmetry restoration temperature $T_c$, one may
obtain a composite Higgs boson $\phi^0_S$ with mass $2m_Q$, a composite
neutral pseudoscalar Nambu-Goldstone boson $\phi^0_P$ and two composite
charged Nambu-Goldstone bosons $\phi^-$ and $\phi^+$.  Thus  the Goldstone
theorem representing the spontaneous breaking of electroweak group
$SU_L(2)\times U_Y(1)\rightarrow U_Q(1)$ is valid rigorously in this case.\\
\indent 2) When one of the two flavors of fermions are massless, we can obtain
the same $\phi^0_S$ and $\phi^0_P$ as ones in 1), but the charged scalar bound
states $\phi^-$ and $\phi^+$ will no longer be rigorously massless and they
could be considered approximately massless only if the momentum cut-off
$\Lambda$ is very large and the considered energy scales are low. This means
that the Goldstone theorem at finite temperature is only approximately valid
at low energy scales.\\
\indent 3) When the two flavors of fermions have unequal nonzero masses, besides
the conclusions in 2) keeping to be valid, it seems that we will meet the possible 
fluctuation effect of the Higgs boson's amplitude originated from the imaginary part 
of its propagator's pole. However, it has been argued that such effect  is  compeletely negligible, if the momentum cut-off $\Lambda$ is sufficiently large.\\
\indent The above conclusions imply that, as far as the discussed
Nambu-Goldstone mechanism in this model is concerned, the two formalisms
of thermal field theory give physically equivalent description.  In particular, by 
the redefinition of  physical propagators in the real-time formalism, the extra 
result of the Higgs mass being doubled which originates from the definition of 
physical propagator of $\phi_S^0$ by 11 element of its matrix propagator  has 
automatically disappeared. \\
\indent Our calculations also show that, the gap equation, which comes from
two-point amputated functions and is determined by the loops bounded by a single
fermion propagator, is identical in the two formalisms  without the need of
analytic continuation.  However, the propagator for scalar bound states , when
defined by four-point amputated functions and calculated in the fermion bubble
diagram approximation, are determined by the loops bounded by two fermion
propagators, will could have some differences in the imaginary parts of their 
denominators in the two formalisms.  Such differences have close relation to the 
introduction of  the ghost fields to cancel the pinch singularities in the 
real-time formalism and  it seems that they could technically be attributed to 
different order of analytic continuation for the discrete Matsubara frequencies in 
the two formalisms [12]. However, a greater possibility is that such differences 
reflect that we are calculating different functions in the two formalisms. This problem needs to be researched further, especially with the bound state propagators being involved in.  Before any definite relation between the 
propagators in the two formalisms is found, we may still respectively use the 
propagators in the two formalisms in the same physical problem and compare the results 
obtained, as has been made in this paper.   Although the differences between the 
propagators do not predict any actually different physical effect in the model 
discussed  here, whether, where and how they could do that is an interesting problem 
deserving of being researched further.
\acknowledgments
This work was partially supported by the National Natural Science Foundation
of China and by Grant No. LWTZ-1298 of the Chinese Academy of Sciences.

\end{document}